\documentclass{pasj00}

\input{pasjsup.sty}

\def\commenta{$^*$}
\def\commentb{$^\dagger$}
\def\commentc{$^\ddagger$}
\def\commentd{$^\S$}

\def\inpress{in press}
\def\astroph#1{ (astro-ph/#1)}

\def\pasjcite2sub#1#2{\authorcite{#1} \yearcite{#1},\yearcite{#2}}
\def\pasjcitet2sub#1#2{\authorcite{#1} (\yearcite{#1},\yearcite{#2})}
\def\pasjcitep2sub#1#2{(\authorcite{#1} \yearcite{#1},\yearcite{#2})}
\def\pasjcite2subyear#1#2#3{\authorcite{#1} \yearcite{#2}#3}
\def\pasjcitet2subyear#1#2#3{\authorcite{#1} (\yearcite{#2}#3)}
\def\pasjcitep2subyear#1#2#3{(\authorcite{#1} \yearcite{#2}#3)}

\DeclareAbbreviation\an{Astron. Nachr.}
\DeclareAbbreviation\AcA{Acta Astron.}
\DeclareAbbreviation\Ap{Astrophysics}
\DeclareAbbreviation\ARep{Astron. Rep.}
\DeclareAbbreviation\ATsir{Astron. Tsirk.}
\DeclareAbbreviation\BaltA{Baltic Astron.}
\DeclareAbbreviation\ibvs{Inf. Bull. Variable Stars}
\DeclareAbbreviation\JAVSO{J. American Assoc. Variable Star Obs.}
\DeclareAbbreviation\JBAA{J. British Astron. Assoc.}
\DeclareAbbreviation\MitVS{Mitteil. Ver\"{a}nderl. Sterne}
\DeclareAbbreviation\MmSAI{Mem. Soc. Astron. Ita.}
\DeclareAbbreviation\Msngr{Messenger}
\DeclareAbbreviation\NewAR{New Astron. Rev.}
\DeclareAbbreviation\Obs{Observatory}
\DeclareAbbreviation\PAZh{Pis'ma AZh}
\DeclareAbbreviation\PZ{Perem. Zvezdy}
\DeclareAbbreviation\PZP{Perem. Zvezdy Pril.}
\DeclareAbbreviation\RMxAA{Rev. Mexicana Astron. Astrof.}
\DeclareAbbreviation\VeSon{Ver\"{o}ff. Sternw. Sonneberg}
\DeclareAbbreviation\VSOLJBul{VSOLJ Variable Star Bull.}
\DeclareAbbreviation\ZA{Z. Astrophys.}

\begin{document}
\SetRunningHead{R. Ishioka et al.}{HV Vir}

\Received{2002/00/00}
\Accepted{2002/00/00}

\title{Period change of Superhumps in the WZ Sge-Type Dwarf Nova, HV Virginis}

\author{%
Ryoko \textsc{Ishioka},\altaffilmark{1}
 Taichi \textsc{Kato},\altaffilmark{1}
 Makoto \textsc{Uemura},\altaffilmark{1}
 Jochen \textsc{Pietz},\altaffilmark{2}
 Tonny \textsc{Vanmunster},\altaffilmark{3}\\
 Tom \textsc{Krajci},\altaffilmark{4}
 Ken'ichi \textsc{Torii},\altaffilmark{5}
 Kenji \textsc{Tanabe},\altaffilmark{6}
 Seiichiro \textsc{Kiyota},\altaffilmark{7}
 Kenzo \textsc{Kinugasa},\altaffilmark{8}\\
 Gianluca \textsc{Masi},\altaffilmark{9}
 Koichi \textsc{Morikawa},\altaffilmark{10}
 Lewis M. \textsc{Cook},\altaffilmark{11}
 Patrick \textsc{Schmeer},\altaffilmark{12}
 Hitoshi \textsc{Yamaoka},\altaffilmark{13}
}
\altaffiltext{1}{Department of Astronomy, Faculty of Science, Kyoto University,
Sakyo-ku, Kyoto 606-8502, Japan}
\email{ishioka@kusastro.kyoto-u.ac.jp, tkato@kusastro.kyoto-u.ac.jp, uemura@kusastro.kyoto-u.ac.jp}
\altaffiltext{2}{Rostocker Str. 62, 50374 Erftstadt, Germany}
\altaffiltext{3}{Center for Backyard Astrophysics (Belgium), Walhostraat 1A, B-3401, Landen, Belgium}
\altaffiltext{4}{1688 Cross Bow Circle, Clovis, New Mexico 88101, USA}
\altaffiltext{5}{Cosmic Radiation Laboratory, Institute of Physical and
Chemical Research (RIKEN), \\
2-1, Wako, Saitama, 351-0198, Japan}
\altaffiltext{6}{Department of Biosphere-Geosphere Systems, Faculty of 
Informatics, Okayama University of Science, \\
Ridaicho 1-1, Okayama 700-0005, Japan}
\altaffiltext{7}{Variable Star Observers League in Japan (VSOLJ);
  Center for Balcony Astrophysics, 1-401-810 Azuma, Tsukuba 305-0031, Japan}
\altaffiltext{8}{Gunma Astronomical Observatory, 6860-86 Nakayama,
Takayama, Agatsuma, Gunma 377-0702, Japan}
\altaffiltext{9}{Physics Department, University of Rome "Tor Vergata",
  Via della Ricerca Scientifica, 100133 Rome, Italy}
\altaffiltext{10}{468-3 Satoyamada, Yakage-cho, Oda-gun, Okayama 714-1213, Japan}
\altaffiltext{11}{Center for Backyard Astrophysics (Concord), 1730 Helix
Ct. Concord, California 94518, USA}
\altaffiltext{12}{Bischmisheim, Am Probstbaum 10, 66132 Saarbr\"{u}cken, 
Germany}
\altaffiltext{13}{Faculty of Science, Kyushu University, Fukuoka 810-8560, Japan}

\newcounter{hvref}
\newcounter{hvrem}



\maketitle

\begin{abstract}
 After 10~years of quiescence, HV Vir underwent a superoutburst in
 January 2002. We report time-series observations clearly revealing the
 period change of ordinary superhumps during the superoutburst. We
 derived a mean superhump period of 0.058260~d and a positive
 period derivative of $7 \times 10^{-5}$. These results are in good
 agreement with the value obtained from the 1992 superoutburst. We also
 detected early superhumps, which were not clearly recognized in the
 past outburst, and a possible rebrightening. Both of them are the
 common characteristics of WZ Sge-type stars.
\end{abstract}

\section{Introduction}

Dwarf novae are a class of cataclysmic variables (CVs), which are close
binaries containing a white-dwarf primary with an accretion disk and a
late-type main-sequence secondary star filling its Roche-lobe
(e.g. \cite{war95book}). They are characterized by repetitive
outbursts. SU UMa stars are a subclass of dwarf novae which show normal
outbursts and superoutbursts. The presence of superhumps during
superoutbursts is a diagnostic of SU UMa stars \citep{war85suuma}. WZ
Sge stars, showing very long recurrence cycle of superoutbursts
(years--decades) and no or few normal outburst during one supercycle,
are extreme SU UMa stars with shortest orbital periods among SU UMa
stars (\cite{bai79wzsge}, \cite{odo91wzsge}). On the contrary, ER UMa
stars or RZ LMi stars are another extreme SU UMa stars with short
orbital periods, which show very short supercycle of a few ten days
(e.g. \cite{kat95eruma}).  About outburst mechanisms of SU UMa stars,
\citet{osa03TTImodel} rejected a recent enhanced mass transfer model (see
\cite{las01DIDNXT}) and proposed a refined tidal-thermal
instability model.

The superhump phenomenon with a period slightly longer than the orbital
period are well explained by the beat of the precession of a tidally
distorted disk with the orbital motion of the secondary
(\cite{whi88tidal}, \cite{hir90SHexcess}). The period of superhumps are
known to decrease during a superoutburst in usual SU UMa stars, and the
typical value of the period derivative is $\sim -5 \times 10^{-5}$
(e.g. \cite{war85suuma}, \cite{pat93vyaqr}). This is attributed to
decreasing apsidal motion due to a shrinking of disk during
superoutburst \citep{osa85SHexcess}. However, several systems are
revealed to show positive period derivations by recent observations
(see, \cite{kat01hvvir}). Most of them are short period systems, such as
WZ Sge stars. 

HV Vir (for a historical review of this object and a recent review of WZ 
Sge stars, see \cite{kat01hvvir}) is one of WZ Sge stars. 
\citet{kat01hvvir} revealed the presence of two superhump periods, the
early superhump\footnote{This feature is also referred to as {\it
orbital superhumps} \citep{kat96alcom}, {\it outburst orbital hump}
\citep{pat98egcnc} or {\it early humps} \citep{osa02wzsgehump}} period of
0.057085(23)~d and the superhump period of 
0.05820(4)~d during the 1992 outburst, and confirmed its WZ Sge
nature. \citet{kat01hvvir} also showed that the period derivative of
superhumps had a positive value of $5.7(6) \times 10^{-5}$, on contrary
to the negative value obtained by \citet{lei94hvvir}.

Most recently, \citet{szk02egcnchvvirHST} reported a temperature of 13300 K for
the white dwarf of HV Vir from the HST observation at 8 yr past the last 
outburst. This temperature is one of the coolest among the white dwarfs
in disk-accreting close binaries, which indicates this system is an
old CV as suggested by other properties \citep{how97periodminimum}. 

After ten years of quiescence since 1992,  a superoutburst of HV Vir was
detected at 2002 January 4.228(UT) (vsnet-alert 6964). We started a
CCD photometric campaign within 0.5 d of the outburst detection. We 
report here on the 2002 outburst of HV Vir and discuss the superoutburst
period change of SU UMa stars.

\section{Observation}

We performed the photometric observations at 11 sites. The journal of
the observations and the equipment used in each site are listed in
table \ref{tab:obs_log}. 

After dark subtraction and flat fielding, we performed aperture
photometry or PSF photometry and obtained differential magnitudes of the
object relative to the comparison stars. The magnitudes obtained from
the unfiltered ST-7E CCD are almost equal to the Rc system.  The magnitude
scales of each observatory were adjusted to that of the Kyoto system 
(comparison star GSC 300.56). 
Heliocentric corrections to observation times were applied
before the following analysis.

\begin{table*}
\caption{Log of the photometric observations.} \label{tab:obs_log}
\begin{center}
\begin{tabular}{rrrccrcc}
\hline\hline
\multicolumn{3}{c}{Date} & HJD$-$2452000 & Exp. Time
 &\multicolumn{1}{c}{N} & Mean Mag. & Instr. $\&$ Comp.$^*$ \\ 
     &    &              & Start -- End                       & (s)
 &                      &           &                       \\
\hline
2002 &January& 4 & 279.194 -- 279.394 &  30 &  623 &13.02(0.04) & A,C$_{a}$,E \\ 
     &    &   & 279.568 -- 279.720 &15-30&  661 &13.09(0.06) & G,H \\   
     &    & 5 & 280.163 -- 280.382 &30-60&  556 &13.21(0.05) & A,C$_{a}$,E \\
     &    &   & 280.973 -- 281.059 &  30 &  218 &13.27(0.04) & J \\
     &    & 6 & 281.171 -- 281.382 &20-60&  678 &13.28(0.07) & A,C$_{a}$,D,E \\
     &    & 7 & 282.282 -- 282.399 &10-60& 1117 &13.40(0.09) & B,D,F \\
     &    &   & 282.957 -- 283.059 &  30 &  255 &13.46(0.06) & J \\
     &    & 8 & 283.168 -- 283.382 &20-60&  664 &13.50(0.08) & A,C$_{b}$,E \\
     &    &   & 283.597 -- 283.780 &  40 &  296 &13.55(0.06) & I \\
     &    &   & 283.953 -- 284.059 &  30 &  268 &13.58(0.06) & J \\
     &    & 9 & 284.170 -- 284.389 &10-60& 2116 &13.61(0.06) & A,B,D,E,F \\
     &    &10 & 285.323 -- 285.338 &  30 &   26 &13.84(0.04) & A \\
     &    &11 & 286.188 -- 286.393 &14-60& 1957 &13.90(0.06) & A,B,E,F \\
     &    &   & 286.546 -- 286.672 &45-60&  114 &13.97(0.05) & H \\
     &    &   & 286.940 -- 287.059 &  30 &  303 &14.01(0.03) & J \\
     &    &12 & 287.228 -- 287.388 &     &  901 &14.02(0.04) & A,B \\
     &    &   & 287.599 -- 287.775 &  50 &  241 &14.13(0.03) & I \\
     &    &13 & 288.259 -- 288.382 &  30 &  247 &14.21(0.06) & C$_{a}$ \\
     &    &   & 288.934 -- 289.059 &  30 &  319 &14.28(0.03) & J \\
     &    &14 & 289.936 -- 290.028 &  30 &  124 &14.35(0.02) & J \\
     &    &15 & 290.981 -- 291.109 &  60 &  143 &14.36(0.06) & K \\
     &    &17 & 292.193 -- 292.326 &  30 &   92 &14.36(0.08) & A \\
     &    &18 & 293.153 -- 293.389 &30-60&  142 &14.4(0.1)  & A,E \\
     &    &   & 293.922 -- 294.056 &  60 &  181 &14.71(0.03) & J \\
     &    &19 & 294.146 -- 294.380 &  30 &  498 &14.7(0.1) & A,C$_{c}$ \\
     &    &20 & 295.926 -- 296.050 & 240 &   44 &16.7(0.1)  & J \\
     &    &21 & 296.244 -- 296.386 &  30 &  311 &17.0(0.4)  & A \\
     &    &22 & 297.133 -- 297.385 &  30 &  739 &17.4(0.6)  & A,C$_{c}$ \\
     &    &23 & 298.132 -- 298.385 &  30 &  489 &17.3(0.5)  & A \\
     &    &24 & 299.145 -- 299.389 &  30 &  833 &17.3(0.4)  & A,C$_{c}$ \\
     &    &   & 299.912 -- 300.046 & 480 &   24 &17.26(0.08)  & J \\
     &    &25 & 300.157 -- 300.375 &  30 &  510 &17.2(0.7)  & A,C$_{c}$ \\
     &    &   & 300.906 -- 301.046 & 480 &   25 &17.35(0.07)  & J \\
     &    &27 & 302.225 -- 302.332 &  30 &  193 &17.4(0.6)  & A \\
     &    &28 & 303.129 -- 303.385 &  30 &  252 &17.3(0.8)  & A \\
     &    &29 & 304.184 -- 304.382 &  30 &  507 &17.5(0.7)  & A,C$_{c}$ \\
     &    &30 & 305.173 -- 305.385 &  30 &  329 &17.5(0.7)  & A \\
     &    &31 & 306.140 -- 306.381 &  30 &  346 &17.3(0.9)  & A \\
     &February& 1 & 307.140 -- 307.316 &  30 &   83 &16.7(0.8)   & A \\
     &    & 3 & 309.261 -- 309.333 &  30 &   34 &17.5(1.1)   & A \\
     &    & 4 & 310.159 -- 310.182 &  30 &   11 &17.4(0.5)   & A \\
     &    &11 & 317.318 -- 317.380 &  30 &   97 &17.8(0.6)  & A \\
2003 &February&14 & 685.348 -- 685.369 &  300&    8 &18.4(0.1) & B\commentb \\ 
\hline
\multicolumn{8}{l}{* A: 30cm tel. + no + SBIG ST-7E, star1 (Kyoto,
 Japan), B: 60cm tel. + R + PixelVision(SITe SI004AB),}\\
\multicolumn{8}{l}{ star1(\commentb star5) (Ouda, Japan), C$_{a}$:  20cm tel. + Apogee
 AP-7p, C$_{b}$: 25cm tel. + Apogee AP-6, C$_{c}$: 25cm tel.}\\
\multicolumn{8}{l}{ + Apogee AP-7p, star1 (Saitama, Japan), D: 30cm
 tel. + no + SBIG ST-9, star1 (Okayama1, Japan), E: 25cm tel.}\\
\multicolumn{8}{l}{ + V + Apogee AP-7,star4 (Tukuba, Japan), F: 25cm
 tel. + no + SBIG ST-7, star2 (Okayama2, Japan), G: 28cm tel. +}\\
\multicolumn{8}{l}{ no + SBIG ST-7, star1 (Ceccano, Italy), H: 20cm + no
 + ST-6B, star1 (Erftstadt, Germany), I: 35cm tel. + no }\\
\multicolumn{8}{l}{ + SBIG ST-7, star1 (Landen, Belgium), J: 28cm tel. +
 no + CB245, star* (New Mexico, USA), K: 44cm tel. + no +}\\ 
\multicolumn{8}{l}{CB245 No.2, star3  (California, USA)} \\
\multicolumn{8}{l}{star1: USNO B1.0 0919-0260839 B1=13.38 R1=12.23}\\
\multicolumn{8}{l}{star2: USNO B1.0 0918-0257674 B1=14.46 R1=13.23}\\
\multicolumn{8}{l}{star3: USNO B1.0 0918-0257733 B1=14.68 R1=13.54}\\
\multicolumn{8}{l}{star4: Tycho-2 300-948 B=11.299 V=10.140 (USNO R1=9.49)}\\
\multicolumn{8}{l}{star5: USNO B1.0 0918-0257808 B1=16.53 R1=15.67}\\
\end{tabular}
\end{center}
\end{table*}

Our spectroscopy was performed on 2002 January 11 with Gunma Compact
Spectrograph (GCS) attached to the classical-Cassegrain focus of the Gunma 
Astronomical Observatory (GAO) 65cm telescope (F/12). We obtained 10
spectra covering 
3800--7670~\AA  with a resolution of 500 with a 5~min exposure time. All
of the raw frames were processed in the usual manner using the IRAF
package\footnote{IRAF is distributed by the National Optical Astronomy
Observatories.}.  

\section{Result}

\subsection{Photometry}

\begin{figure}
 \begin{center}
  \FigureFile(88mm,115mm){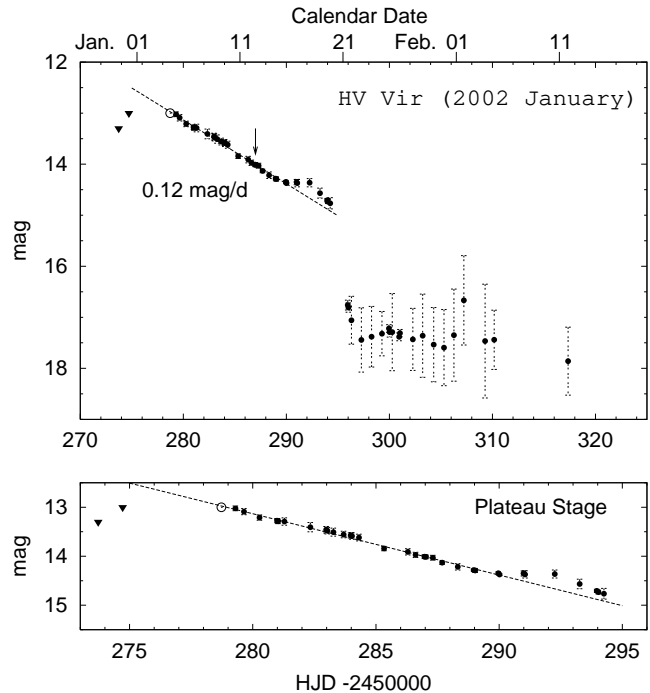}
 \end{center}
 \caption{ Superoutburst in 2002 January.  The open circles and filled
 triangles are the outburst detection and last negative visual observations
 reported to VSNET, respectively.  The dots are our CCD observations.  
 The bars on dots represent one sigma calculated using data on eah
 night. The dashed line represents the linear fit for the
 gradually declining plateau stage. The arrow indicated the time of
 spectroscopy. Upper panel: The whole light curve of the outburst. Lower panel: 
 Enlarged light curve of the plateau stage.}
 \label{fig:longlc}
\end{figure}

The whole light curve of the superoutburst in 2002 is shown in figure
\ref{fig:longlc}. The outburst was detected at $m_{\rm vis}=$13.3 on 2002
January 4.228 (UT, HJD~2452278.728). The last negative observation was
reported on 2001 December 31.219 (UT, HJD 2452274.718). Our CCD
observations started 0.5~day after the outburst detection. The object
faded at a constant rate of 0.12~mag~d$^{-1}$ for 11 days. During the
following four days, the gradual decline stopped and the object remained
at an almost constant magnitude around $R_{\rm c}=$14.4~mag. The main
superoutburst lasted at least for 14~days until the rapid decline
started on January 18. The object declined by 3~mag to 17.4~mag within
3~days. 
The light curve was constant within their one sigma, after the
rapid decline ended. However the average magnitude on 2002  February 1 was
brighter by $\sim 1$~mag than those of other nights. 
It may be a rebrightening, although it is a
difficult thing to really know given the data in hand.
One year later, the object has become 0.5~mag fainter than just after
the outburst. 

\begin{figure}
 \begin{center}
   \FigureFile(88mm,115mm){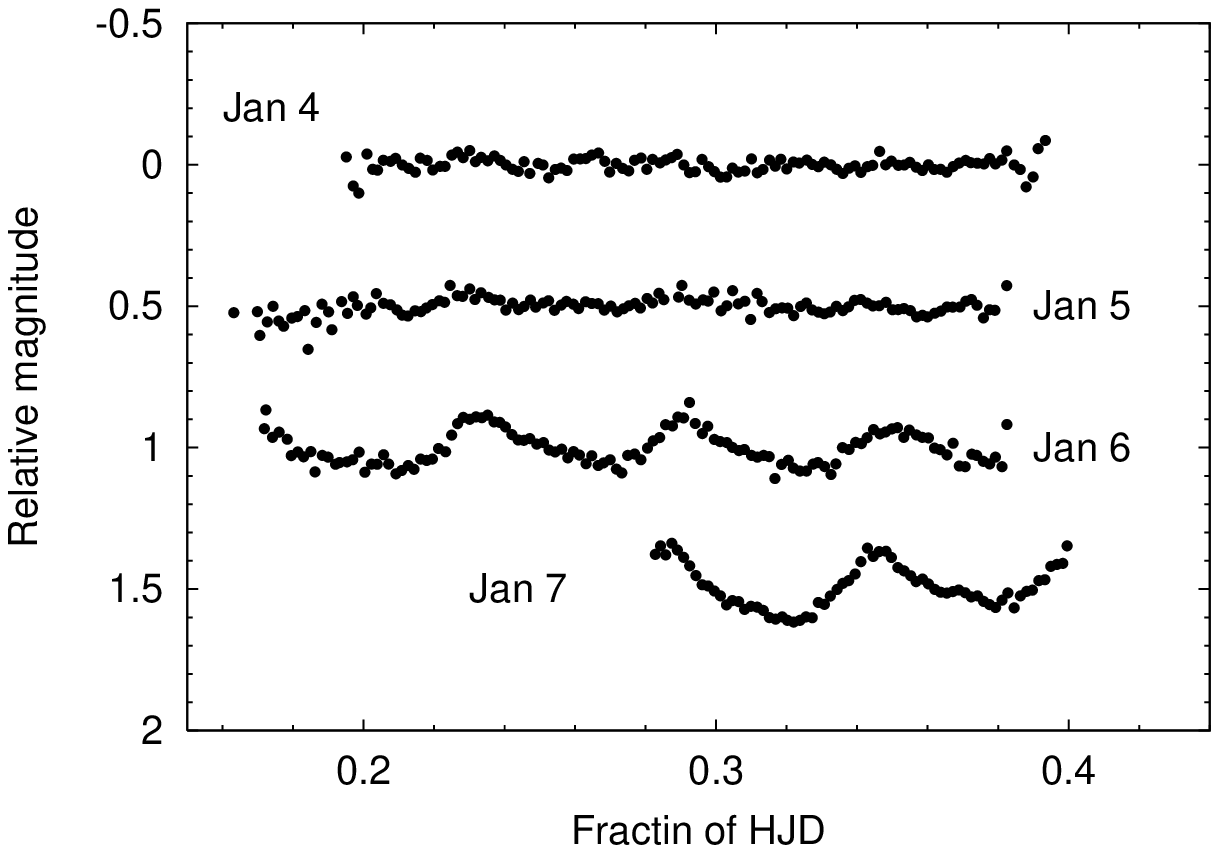}
 \end{center}
 \caption{Enlarged light curves in first four days.  Each point
 represents the average of 0.02~d bin with the typical standard error
 of 0.005--0.02~mag. Humps showing steeper rise and a slower decline
 appeared in the third day.}
 \label{fig:daylc}
\end{figure}

\begin{figure}
 \begin{center}
   \FigureFile(88mm,115mm){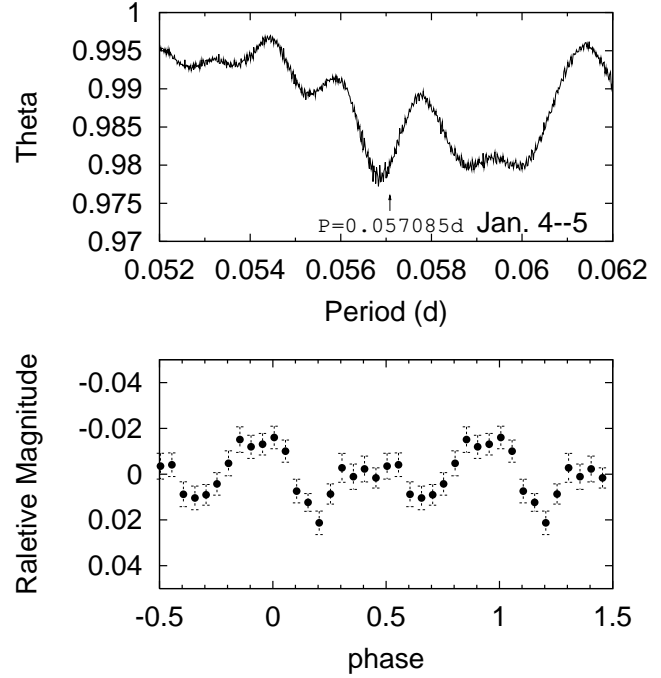}
 \end{center}
 \caption{(Upper) Period-theta diagram of the PDM period analysis on
 the data during the first two days. The abscissa and ordinate are the
 period in days and the theta, respectively. A weak signal appears
 around a period of 0.0569(1)~d. This period is close to the early
 superhump period of 0.057085~d obtained for the 1992 superoutburst by
 Kato et al. 2001 (shown by an arrow in the figure). (Lower) Phase-averaged
 light curve with a period of 0.057085~d.} 
 \label{fig:early}
\end{figure}

Figure \ref{fig:daylc} shows the nightly light curves of the first four
days. The light curves on January 4 and 5 show complex modulation with
small amplitudes. We performed a period analysis using the Phase
Dispersion Minimization (PDM) method \citep{PDM} to the data on these
two days, after removing the linear trend of the decline. The resultant
period-theta diagram shows a very weak signal with a period of 0.0569(1)~d
(upper panel of figure \ref{fig:early}). This period is close to the
early superhump period of 0.057085(23)~d obtained by
\cite{kat01hvvir}. The lower panel of figure \ref{fig:early} shows
the averaged light curve of the January 4 and 5 data folded by the
period of 0.057085~d. The averaged light curve has a double peaked
feature with an amplitude of 0.04~mag.

\begin{table}
\caption{Timings of superhumps.} \label{tab:superhump_time}
\begin{center}
\begin{tabular}{cc||cc}
\hline
HJD$-$2452000     &  E& HJD$-$2452000    &  E \\
\hline
280.9980 &-22& 286.9965 & 81\\
281.2325 &-18& 287.0520 & 82\\
281.2905 &-17& 287.2860 & 86\\ 
281.3495 &-16& 287.3450 & 87\\ 
282.2880 &  0& 287.6370 & 92\\
282.3449 &  1& 287.6980 & 93\\ 
282.9820 & 12& 287.7575 & 94\\ 
283.0410 & 13& 288.2815 &103\\ 
283.2145 & 16& 288.3380 &104\\ 
283.2740 & 17& 288.9830 &115\\ 
283.3290 & 18& 289.0400 &116\\ 
283.6200 & 23& 289.9770 &132\\ 
283.6800 & 24& 291.0270 &150\\ 
283.7370 & 25& 291.0860 &151\\ 
283.9670 & 29& 293.1780 &187\\
284.0255 & 30& 293.2320 &188\\
284.2020 & 33& 293.2910 &189\\
284.2580 & 34& 293.9300 &200\\
284.3185 & 35& 293.9890 &201\\ 
284.3760 & 36& 294.0430 &202\\ 
286.2370 & 68& 294.2190 &205\\ 
286.2950 & 69& 294.2760 &206\\
286.3505 & 70& 294.3320 &207\\
\hline	       
\end{tabular}  
\end{center}
\end{table}

On January 6, another variability with a large amplitude appeared
(figure \ref{fig:daylc}). 
We performed a PDM analysis on the data during
January 6--18, after subtraction of the decline trend. 
A strong signal exists at the period of 0.058217(12)~d.

We measured the maximum times of the superhumps mainly by eye but
when it was difficult to determine an accurate maximum, we used a
cross-correlation method to obtain individual maxima. The
errors of the maximum times are 0.001--0.005~d. All the peak times we
obtained are listed in table \ref{tab:superhump_time}. The values are
given to 0.0001~d, though the error is larger than 0.001~d, 
in order to avoid any loss of significant digits in a later
analysis. The cycle count ($E$) denotes the cycle number of humps
counted from HJD 2452282.2880. A linear ephemeris for the peaks is
represented by the following equation: 
\begin{equation}
{\rm HJD_{max}} = 2452282.2814(10) + 0.058238(8) E. \label{equ:1}
\end{equation}

\begin{figure}
 \begin{center}
   \FigureFile(88mm,115mm){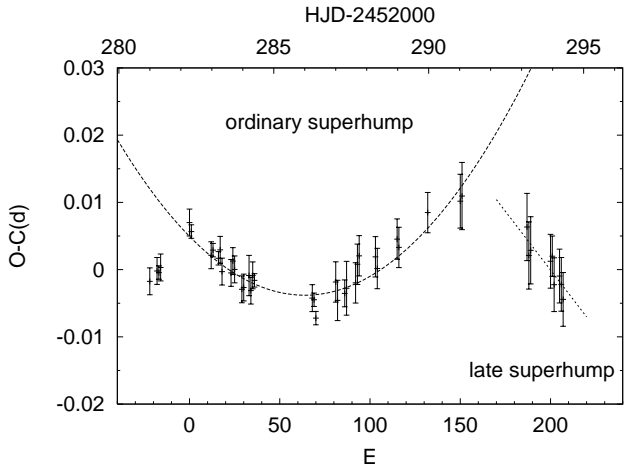}
 \end{center}
 \caption{O-C diagram of superhumps. The parabolic solid line
 corresponds to the quadratic ephemeris represented by equation (2). 
 }
 \label{fig:oc}
\end{figure}

\begin{figure}
 \begin{center}
   \FigureFile(88mm,115mm){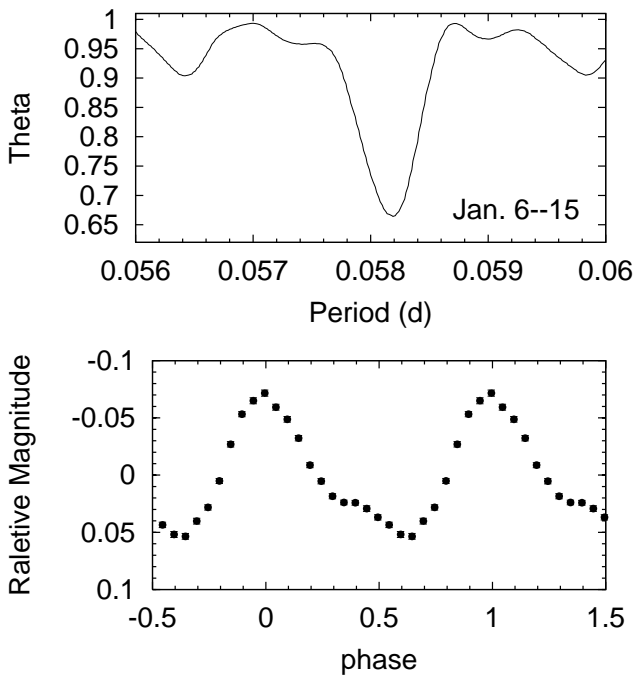}
 \end{center}
 \caption{(Upper) Period-theta diagram of the PDM period analysis on
 the data during HJD 2452282--2452291. The abscissa and ordinate are the
 period in days and the theta, respectively. (Lower) Phase-averaged
 light curve with a period of 0.0582034(6)~d.} 
 \label{fig:super}
\end{figure}

The $O-C$s calculated for Eq. \ref{equ:1} is displayed in figure
\ref{fig:oc}. 
The diagram indicates that the superhumps were consinted by three types
of humps with cycle counts of $E=-22\sim0$, $E=0\sim151$, and $E=170\sim220$.

The times of peaks with cycle counts of $E=-22\sim0$ (corresponding on
January 6) 
were slightly earlier than expected from the quadratic 
ephemeris of following peaks. We made a PDM period analysis to the
January 6 data, and 
obtained a period of 0.05844(24)~d. Although the error is large, this
period is significantly longer than the period of 0.058026(31)~d for the
next three days.

The $O-C$s for superhumps during gradually declining part of the plateau stage 
(HJD~2452282--2452291) clearly show that the superhump period
increased. The best fit quadratic ephemeris is represented by the
following equation:  
\begin{eqnarray}
{\rm HJD_{max}} = 2452882.2866(7) + 0.057949(26) E \nonumber \\
       +2.2(2) \times 10^{-6} E^{2}. \label{equ:2}
\end{eqnarray}
The rate of the period change is $P_{\rm dot}=\dot{P}_{SH}\}/P_{\rm SH}=7.8(7)
\times 10^{-5}$. 
From the PDM analysis to the data sets of this stage, a mean superhump
period of 0.0581836(30)~d was yielded. The theta
diagram and the phase-averaged light curve with this period are
exhibited in figure \ref{fig:super}.

It is well known that humps called late superhumps, which have a period close
to the superhump period but show a 0.5 phase jump from that of ordinary
superhumps, appear during the very late stage of, or shortly after a
superoutburst (\cite{hae79lateSH}; \cite{vog83lateSH};
\cite{vanderwoe88lateSH}; \cite{hes92lateSH}).

The $O-C$ diagram also shows that the timings of the humps  during
January 18--20 ($E>170$) deviated from the expected line for ordinary
superhumps during the plateau stage. From the linear
regression to the peak times, we obtained a late superhump period of
0.057889(78)~d corresponding to the fitted straight line in figure
\ref{fig:oc}, and the PDM analysis for the January 17 and 20 data gave
a period of 0.0578853(25)~d.

\begin{figure}
 \begin{center}
   \FigureFile(88mm,115mm){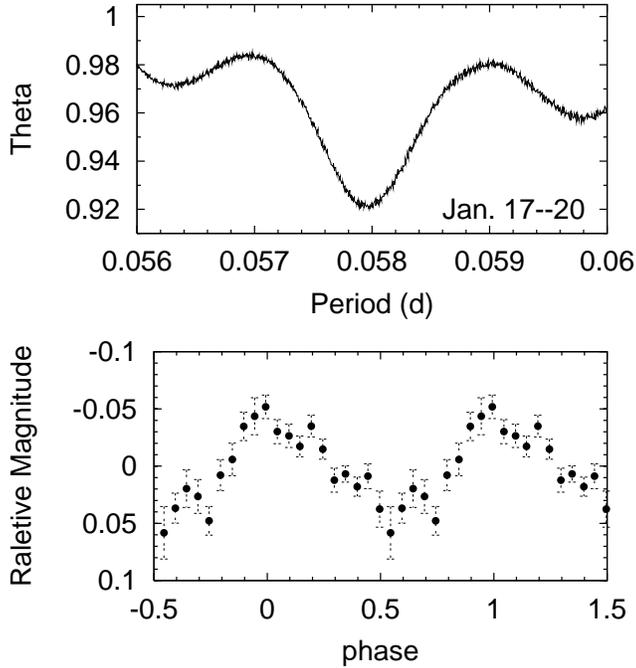}
 \end{center}
 \caption{(Upper) Period-theta diagram of the PDM period analysis on
 the data during HJD 2452293--2452296. The abscissa and ordinate are the
 period in days and the theta, respectively. (Lower) Phase-averaged
 light curve with a period of 0.0578853(25)~d.} 
 \label{fig:late}
\end{figure}

The amplitude of superhumps reached a maximum (0.2~mag) at around
January 7--8, and gradually decayed to 0.08~mag on January 17.
On the contrary, secondary maxima around the phase 0.4, which appeared at
January 11, kept an almost constant amplitude of $\sim$ 0.05~mag.

After the rapid fading stage, we could not detect any periodic modulations due
to the faintness of the object.

\subsection{Spectroscopy}

\begin{figure}
 \begin{center}
   \FigureFile(88mm,115mm){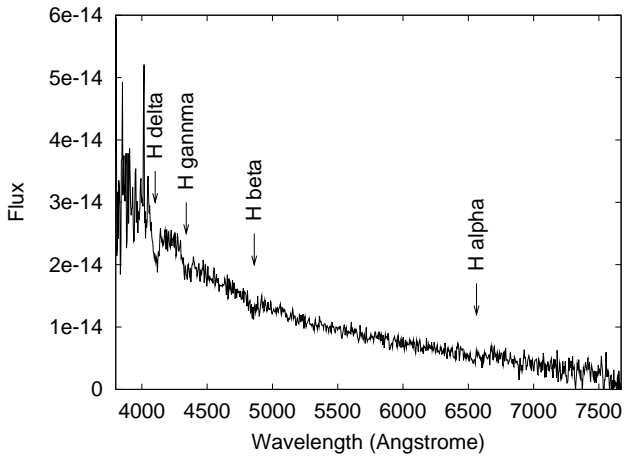}
 \end{center}
 \caption{Spectrum during the outburst. Balmer lines are in
 absorption. Flux units are erg cm$^{-2}$ c$^{-1}$ \AA$^{-1}$.}
 \label{fig:spec}
\end{figure}

Figure \ref{fig:spec} shows the flux-calibrated mean spectrum. Balmer lines in
absorption superimposed on a blue continuum. The H$\beta$ absorption line
is clearly detected, however, H$\alpha$ is not seen. This spectrum is
similar to the spectrum of AL Com during the outburst plateau stage
(\cite{szk96alcomIUE}, \cite{kil92hvviriauc}). 

\section{Discussion}

\subsection{Astrometry and Identification}

\begin{table}
\caption{Astrometry of HV Vir.}\label{tab:astrometry}
\begin{center}
\vspace{10pt}
\begin{tabular}{cccc}
\hline\hline
Source    & R. A. & Decl.               & Epoch \\
          & \multicolumn{2}{c}{(J2000.0)} & \\
\hline
USNO A2.0  &13 21 03.133 & +01 53 29.65 & 1956.270  \\
GSC 2.2    &13 21 03.174 & +01 53 29.29 & 1996.353  \\
USNO B1.0  &13 21 03.173 & +01 53 29.01 & 2000.0    \\
\hline
\end{tabular}
\end{center}
\end{table}

  Using outburst images, we measure the position of HV Vir as
R.A. = 13$^{\rm h}$ 21$^{\rm m}$ 03$^{\rm s}$.174, Decl. = +01$^\circ$
53$\prime$ 29$\prime\prime$.63 (J2000.0, USNO-A2.0 frame).  The
mean error is about 0''.2 for each coordinate.  It confirms the
quiesient counterpart identification with an USNO-A2.0 star (position
end figures 03$^{\rm s}$.133, 29$\prime\prime$.65, red mag = 19.0, blue
mag = 18.7, Epoch = 1956.270).  From these values, we estimate the
proper motion of HV Vir as about 13 $\pm$ 8 mas/yr, which is
consistent with (though somewhat smaller than) the USNO-B1.0 value of
23 $\pm$ 9 mas/yr.  

  If we assume the distance of HV Vir \citep{szk02egcnchvvirHST} six
times as large as that of WZ Sge \citep{dhi00CVIRspec}, 
the transversal velocity of HV Vir is estimated as 33 km/s.  It
is comparable with that of WZ Sge (40 km/s).

\subsection{Supercycle}

Three outbursts of HV Vir were detected on Sonneberg plates, respectively at 
13.5 mag in 1939, 11.5 mag in 1970 and 14.0 mag in 1981 \citep{lei94hvvir}.
The 2002 superoutburst occurred after 10 years of quiescence from the 1992
superoutburst. Since 1992, no outburst was detected till the 2002
outburst.\footnote{Before the 2002 outburst, two unconfirmed positive
detections were reported to AAVSO at 14.2 mag on 52013 and 13 mag on
52055. However, no positive detection was reported to VSNET during this
period.} \citet{lei94hvvir} and \citet{kat01hvvir} respectively implied the
recurrence cycle to be $\sim$ 10 years or shorter and to be $\sim$ 10
years or even longer. Our detection of the 2002 superoutburst confirmed
that he recurrence cycle of superoutbursts was $\sim$~10 years. The long
recurrence cycle of superoutbursts and the absence 
of normal outbursts are common properties of the WZ Sge-type dwarf novae.

\subsection{Outburst Light Curve}

During the 1992 superoutburst, the visual maximum was reached at $V=11.5$
\citep{kil92hvviriauc}. The superhumps appeared 7 to 9~d after the
visual maximum and the main outburst lasted for 23 d. 
Our detection of the 2002 superoutburst was at $m_{\rm vis}=13.3$. The delay of
superhumps was $<$~6~d and the main outburst lasted for $<$~20~d. If we
assume the superoutburst maximum was within 0.5 d after the last 
negative observation and the declining rate of the first stage was the same
as that of the 1992 outburst, the visual maximum of the 2002 outburst
is estimated to be $\sim$~12.0 mag. 
The declining rate during the first 10~d of the plateau stage was almost
constant at 0.12 mag d$^{-1}$. During the last 4~d, the declining
stopped and the light curve became flat. The declining rate of the
outburst plateau of WZ Sge stars, decrease from 0.2 mag d$^{-1}$ during the
first few days to 0.13 mag d$^{-1}$, a typical value for SU UMa stars.
We missed the first rapid declining stage. 

The outburst maximum and the duration show that the 2002 superoutburst
of HV Vir was of a slightly smaller scale than the 1992 outburst. This is 
consistent with the quiescence durations of 11~yr and 10~yr before
respective superoutbursts, since the mass stored in the disk before the
1992 outburst was estimated to be slightly more than the 2002 outburst
if the mass transfer rate was constant.

Our observations also covered the two weeks after the fading from the 
main outburst, which were not well observed in the 1992 outburst. 
The average magnitude on the 11th day after the end of the rapid decline
was brighter than those of other nights. If it indicates a
rebrightening, the 10-days quiescence 
before the rebrightening is rather long compared with other WZ Sge stars
or SU UMa stars with rebrightening(s). For example, in EG Cnc this was
one week, and in WZ Sge only 2 or 3 days. The mass amount stored in the
accretion disk increase  with the length of the quiescence period, so
the amplitude of the rebrightening is expected to be larger as well. 
Unfortunately, the error bar in our observations is fairly
large. and we were unable to obtain data on the following 2 nights.  
Therefore, we cannot know about  the amplitude and can only state that
the outburst duration of the rebrightening was shorter than 3~days.

\subsection{Humps During the Superoutburst}

As described in section 3.1, during the first two days of our
observation, we obtained small-amplitude humps with a shorter period
than the period of the following days, which is close to the early superhump
period ($P_{\rm e}=0.057085$~d) derived by \citet{kat01hvvir}. The averaged 
light curve folded by $P_{\rm e}$ shown in figure \ref{fig:early} has a
double-peaked profile that is common to early superhumps of WZ Sge
stars. Thus we can conclude that the humps observed in the first two 
days are early superhumps just before decaying. 
 
The large-amplitude humps on the third day of our observation shows something
different on the $O-C$ diagram (figure \ref{fig:oc}) from the following
ordinary superhumps. The PDM analysis indicate that the period of four
humps with cycle counts from $-22$ to $-16$ are longer than on following days. 
These humps may be a transition state of early superhumps and ordinary 
superhumps. The transition occurred within 2 days. However, due to
the 1~d gap of our observations, the precise length of the transition
state was not determined and it is not clear whether the phase jump existed or not.
A similar behavior of $O-C$ was also observed during the 2001 superoutburst of
WZ Sge \citep{ish02wzsgeletter}. The transition between early superhumps
and ordinary superhumps occurred in 0.5 day and the period of humps
during the transition was also longer than following days. In the case
of WZ Sge, the phase jump did not exist.

The mean period of ordinary superhumps, 0.0581836(30) d, and the period
derivative, $7(1)\times10^{-1}$ in the 2002 outburst, are in good
agreement with those of the 1992 outburst derived by \citet{kat01hvvir}. 

The humps with a period of  0.05788(7)~d during the last stage of the
superoutburst are very likely to 
be late superhumps. Due to the 2 days gap of
peak timimngs, it is unclear whether the 0.5 phase jump really occured or not. 
We could not detect any modulation after the rapid declining. However,
the late superhump period we obtained is very close to the period of
0.05799(3)~d which \citet{lei94hvvir} obtained for the low state up to
$\sim$20~days after the rapid declinig of the 1994 superoutburst of HV
Vir and identified with  the orbital period. This confirmes that the
period of 0.05799(3)~d is identified with late superhumps as seggested by
\citet{kat01hvvir}, and late superhumps in superoutbursts of HV Vir persist
for a long time with a constant period as in other WZ Sge stars
(\cite{kat97tleo}; \cite{pat98egcnc}).

\subsection{Superhump Period Change} 

\begin{table}
\caption{Superhump period change.}\label{tab:pdot}
\setcounter{hvref}{0}
\setcounter{hvrem}{0}
\begin{center}
\begin{tabular}{llcc}
\hline\hline
Object\commenta & $P_{\rm SH}$\commentb & $P_{\rm dot}$\commentc & Ref. \\
\hline
J2329         & 0.046271  & 11.9(2.4)   & \ref{count:pdot:j2329-1} \\
HV Vir        & 0.058203  & 7.8(7)        & \ref{count:pdot:hvvir-1} \\
CC Cnc        & 0.0755135 & $-$10.2(1.3)& \ref{count:pdot:cccnc-1} \\
V877 Ara      & 0.08411   & $-$14.5(2.1)& \ref{count:pdot:v877ara-1} \\
EF Peg(1997)  & 0.0871    & 0(3) \commentd & \ref{count:pdot:efpeg-1} \\
BF Ara        & 0.08797   & $-$0.8(1.4) & \ref{count:pdot:bfara-1} \\
KK Tel        & 0.08801   & $-$37(4)    & \ref{count:pdot:kktel-1} \\
V725 Aql      & 0.09909   & 0(3) \commentb & \ref{count:pdot:v725aql-1} \\
\hline
 \multicolumn{4}{l}{\commenta Year of the outburst in parentheses.} \\
 \multicolumn{4}{l}{\commentb Typical superhump period (d).} \\
 \multicolumn{4}{l}{\commentc $\dot{P}/P$ unit in 10$^{-5}$.} \\
\end{tabular}
\end{center}
{\footnotesize  
  {\bf Remarks:}
   \refstepcounter{hvrem}\label{rem:pdot:v725aql-1}
    \arabic{hvrem},
   \refstepcounter{hvrem}\label{rem:pdot:efpeg-1}
    \arabic{hvrem}. Zero or marginally positive $P_{\rm dot}$ is indicated,
    although definite value is not given. 

  {\bf References:}
   \refstepcounter{hvref}\label{count:pdot:j2329-1}
    \arabic{hvref}. \citet{uem02j2329} ;
   \refstepcounter{hvref}\label{count:pdot:hvvir-1}
    \arabic{hvref}. this paper ;
   \refstepcounter{hvref}\label{count:pdot:cccnc-1}
    \arabic{hvref}. \citet{kat02cccnc} ;
   \refstepcounter{hvref}\label{count:pdot:v877ara-1}
    \arabic{hvref}. \citet{kat02v877arakktelpucma} ;
   \refstepcounter{hvref}\label{count:pdot:v725aql-1}
    \arabic{hvref}. \citet{uem01v725aql} ;
   \refstepcounter{hvref}\label{count:pdot:bfara-1}
    \arabic{hvref}. \citet{kat02bfara} ;
   \refstepcounter{hvref}\label{count:pdot:efpeg-1}
    \arabic{hvref}. K. Matsumot at al. inpreparation ;
   \refstepcounter{hvref}\label{count:pdot:kktel-1}
    \arabic{hvref}. \citet{kat02v877arakktelpucma} ;
 }
\end{table}

As noted in section 1, the superhump periods in usual SU UMa stars decrease
during a superoutburst and it is explained by the shrinking of the accretion
disk. However, several systems with short periods, 
including WZ Sge stars, show positive period derivatives such as HV Vir.
\citet{kat01hvvir} made a systematic survey of the $P_{\rm dot}$ in SU UMa
stars and found that short-period or infrequently outbursting systems
show an increase of the superhump periods. They suggest that low $\dot{M}$
is more reasonable, because some long period systems with low $\dot{M}$ also
show positive $P_{\rm dot}$. For the explanation of positive $P_{\rm dot}$, it is
proposed that the eccentricity can propagate outward and the superhump
period increases when the system has an expanding disk beyond the 3:1
resonance radius because of a large amount of accumulated matter due to
a small mass transfer rate and/or a small 3:1 resonance radius due to a
small mass ratio. 

\begin{figure}
 \begin{center}
   \FigureFile(88mm,115mm){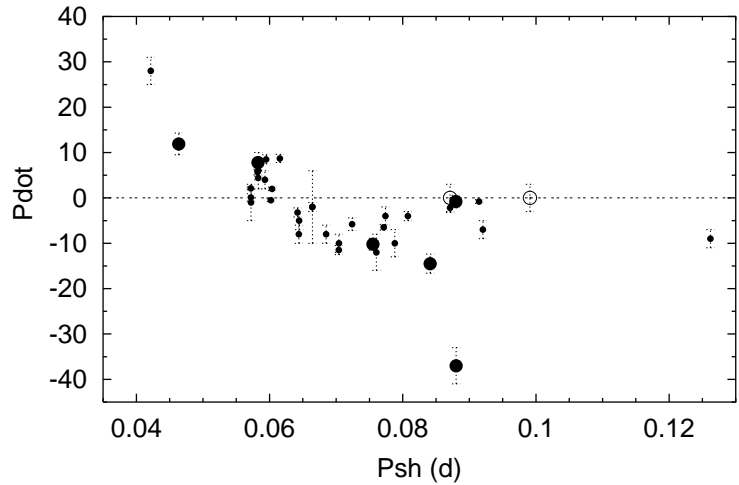}
 \end{center}
 \caption{$P_{\rm dot}=\dot{P}_{\rm SH}/P_{\rm SH}$ versus $P_{\rm SH}$ diagram. The data of table 3 are
 added to the original figure (figure 10. in Kato et al. (2002)).
 }
 \label{fig:pdot}
\end{figure}

We added some new points listed in table \ref{tab:pdot} to figure 10 of
\citet{kat01hvvir} (figure \ref{fig:pdot}). 1RXS J232953.9$+$062814 is
one of the shortest period systems and shows a large superhump period
increase. Two long-period SU UMa stars, V877 Ara and KK Tel show
large decreases of the superhump period \citep{kat02v877arakktelpucma},
however BF Ara with a similar period shows an almost zero or marginally
positive period change \citep{kat02bfara}. V725 Aql \citep{uem01v725aql}
and EF Peg during the 1996 outburst (K. Matsumoto et al., in preparation)
are also indicated zero or marginally positive $P_{\rm dot}$. 

Figure \ref{fig:pdot} shows that $P_{\rm dot}$ is positive in short-period
systems, and a minimum exists around superhump periods of 0.07--0.08~d,
except for V877 Ara and KK Tel. 

This diagram cannot explain only through the above scenario, that a positive
$P_{\rm dot}$ is the result of a small mass transfer rate. 
1 RXS J232953.9$+$062814 shows a large $P_{\rm dot}$, although a high
$\dot{M}$ is indicated for this stars \citep{uem02j2329letter}. 
BF Ara is another system which shows inconsistency with the above
scenario. This object shows rather frequent outbursts indicating a
large mass transfer rate, however  a virtually zero $P_{\rm dot}$ is observed, in
contrary to the similar systems of V877 Ara and KK Tel, which show a very
large decrease of superhump periods. 

These systems indicate that the values of $P_{\rm dot}$ cannot be explained by one 
parameter of period, mass ratio, or mass transfer rate, and that if the
positive $P_{\rm dot}$ is a result of expansion of the disk beyond the 3:1
resonance radius, the cause of expansion is not only the low mass ratio, 
and/or low mass transfer rate, but another mechanism is needed.

In figure \ref{fig:pdot}, the period derivatives of ER UMa stars,
another extreme sub-group of SU UMa stars with a short period and a high
mass transfer rate, are not included except for V1159 Ori which has a
typical $P_{\rm dot}$. The examination of $P_{\rm dot}$ of other
ER UMa stars, especially the systems with shorter periods such as RZ LMi 
and DI UMa, may give some constraint on the mechanism for variation of
$P_{\rm dot}$ in SU UMa stars.  

\section{Conclusion}

Our photometric observation of the 2001 outburst in HV Vir confirmed the
presence of early superhumps and the positive derivative of the
superhump period.

We revealed that the humps during the transition from early
superhumps to ordinary superhumps have a period sulightly longer than
mean superhump period, as observed during the 2001 superoutburst in WZ Sge.

Our observations also revealed that at the late stage of the
superoutburst, the late superhumps appeared and they persist more than
20~d after the main outburst terminates.
After the main outburst, one possible rebrightening was detected.  
The long quiescent state before the rebrightening suggests that the
rebrightening had a large amplitude but due to the lack of observations, 
this was not confirmed.  

It is clear that the derivatives of superhump periods of WZ Sge stars
are positive, but those of SU UMa stars, including WZ Sge stars, are not
simply explained by 
their orbital periods or mass transfer rates. For the systematic study,
more observations, especially of long period systems and systems
with a short period and a large mass transfer rate, are needed.

\vskip 3mm

We are grateful to many amateur observers for supplying their vital
visual CCD estimates via VSNET and AAVSO.  Part of this work is
supported by a Research Fellowship of the Japan Society for the
Promotion of Science for Young Scientists (MU).

\end{document}